\documentclass[twocolumn,amsmath,amssymb,prb,superscriptaddress]{revtex4-1}

\usepackage{titlesec}
\usepackage[colorlinks,bookmarks=false,citecolor=blue,linkcolor=red,urlcolor=blue]{hyperref}
\usepackage{epsfig}
\usepackage{color}
\usepackage{subfigure}
\usepackage{graphicx}    
\usepackage{dcolumn}     
\usepackage{lipsum}      
\usepackage{multirow}
\usepackage{nicefrac}

\usepackage{natbib}
\setcitestyle{square,numbers}

\input{insbox}

\newcommand{\be}{\begin{equation}}
\newcommand{\ee}{\end{equation}}

\definecolor{drkgr}{rgb}{0.05,0.6,0.2}

\definecolor{Ora}{cmyk}{0, 0.6, 0.8, 0.4}

\definecolor{pink}{cmyk}{0.21, 0.74, 0, 0}

\definecolor{mygray}{cmyk}{0, 0, 0, 0.3}
\usepackage[normalem]{ulem}

\newcommand{\ICVO}{InCu$_{2/3}$V$_{1/3}$O$_3$}

\begin{document}

\title{Interplay of electron correlations, spin-orbit couplings, and structural effects for Cu
       centers in the quasi-two-dimensional magnet \ICVO}

\author{R.~Murugesan}
\affiliation{Leibniz IFW Dresden, Helmholtzstr.~20, 01069 Dresden, Germany}

\author{M.~S.~Eldeeb}
\affiliation{Leibniz IFW Dresden, Helmholtzstr.~20, 01069 Dresden, Germany}

\author{M.~Yehia}
\affiliation{Leibniz IFW Dresden, Helmholtzstr.~20, 01069 Dresden, Germany}
\affiliation{Reactor Physics Department, Nuclear Research Center, Atomic Energy Authority, Cairo, Egypt}


\author{B.~B\"{u}chner}
\affiliation{Leibniz IFW Dresden, Helmholtzstr.~20, 01069 Dresden, Germany}
\affiliation{Institute for Solid State and Materials Physics and W{\"u}rzburg-Dresden Cluster of Excellence ct.qmat, TU Dresden, D-01062 Dresden, Germany}

\author{V.~Kataev}
\affiliation{Leibniz IFW Dresden, Helmholtzstr.~20, 01069 Dresden, Germany}

\author{O.~Janson}
\affiliation{Leibniz IFW Dresden, Helmholtzstr.~20, 01069 Dresden, Germany}

\author{L.~Hozoi}
\affiliation{Leibniz IFW Dresden, Helmholtzstr.~20, 01069 Dresden, Germany}

\vspace{2.0cm}

\begin{abstract}
\noindent
Less common ligand coordination of transition-metal centers is often associated with peculiar valence-shell
electron configurations and outstanding physical properties.
One example is the Fe$^+$ ion with linear coordination, actively investigated in the research area of
single-molecule magnetism.
Here we address the nature of 3$d^9$ states for Cu$^{2+}$ ions sitting in the center of trigonal bipyramidal
ligand cages in the quasi-two-dimensional honeycomb compound \ICVO, whose unusual magnetic
properties were intensively studied in the recent past.
In particular, we discuss the interplay of structural effects, electron correlations, and spin-orbit
couplings in this material.
A relevant computational finding is a different sequence of the Cu ($xz$,\,$yz$) and ($xy$,\,$x^2\!-\!y^2$)
levels as compared to existing electronic-structure models, which has implications for the interpretation of
various excitation spectra.
Spin-orbit interactions, both first- and second-order, turn out to be stronger than previously assumed,
suggesting that rather rich single-ion magnetic properties can be in principle achieved also for the 3$d^9$
configuration by properly adjusting the sequence of crystal-field states for such less usual ligand
coordination.
\end{abstract}

\date\today
\maketitle

\section{Introduction}

Transition-metal (TM) ions with atypical ligand (L) coordination in TM pnictides, chalcogenides, and halides
may display quite unexpected features as concerns their electronic structures and physical properties.
A remarkable example is monovalent Fe with linear L-Fe-L bonds, which due to its impressively strong magnetic
anisotropy is being presently investigated in the context of single-molecule magnetism \cite{fe_d7_Zadrozny_2013,
fe_d7_Jesche_2014}.
Also rare is for instance the square coordination realized for nickel (formally 3$d^9$) in NdNiO$_2$.
Interestingly, superconductivity with $T_{\mathrm c}$'s of up to 15 K was recently reported in this system
\cite{nio2_li_19}, the available data suggesting physics significantly different as compared to the Cu-oxide
superconductors \cite{nio2_hepting_19,nio2_berciu_19}.

The title compound of the present study, honeycomb \ICVO, has attracted significant attention due to
its peculiar magnetic properties \cite{Kataev2004,Moeller2008,Yehia2010,Okubo2011,Yan2012,Jia2017,Jia2017a}.
In particular, it has been found recently \cite{Iakovleva2019} that a pronounced two-dimensional (2D)
$XY$ anisotropy of the Cu-based honeycomb spin-1/2 planes and strong inter-layer frustration yield an
extended low-temperature regime where a quasi-2D static magnetic state is stabilized, showing some
indications of a Berezinsky-Kosterlitz-Thouless transition to the topologically ordered state of
vortex-antivortex pairs predicted in the 2D {\it{XY}} model
\cite{Berezinskii1971,Kosterlitz1972,Kosterlitz1973,Kosterlitz1974}. 
An interesting aspect is the uncommon type of ligand environment encountered in this compound, with
trigonal ligand bipyramids around each copper ion.
For $D_{3h}$ Cu-site symmetry as realized in this situation, there are two sets of doubly degenerate
orbitals ($xz$, $yz$ and $xy$, $x^2\!-\!y^2$), as for magnetically anisotropic L-TM-L units with either
$D_{6h}$ \cite{fe_d7_Jesche_2014,fe_d7_xu_2017,fe_d7_Jesche_2015,fe_d7_Antropov_2014} or approximate
$D_{3h}$ \cite{fe_d7_Zadrozny_2013} symmetry.
Finding that spin-orbit (SO) interactions play an important role for such hole configurations in various
L-TM-L complexes motivates then a detailed investigation of SO interactions for Cu 3$d^9$ hole states 
in \ICVO.
Even if the $z^2$ ground-state character inferred from analysis of electron spin resonance (ESR) spectra 
\cite{Kataev2004} and by band-structure calculations \cite{InCuVO_lado_pardo_16,Iakovleva2019} indicates
that to first order only the excited states would be affected by SO couplings in this particular cuprate,
such a study should still provide insights into the prospect of strong single-ion magnetic anisotropy
for 3$d^9$ centers with suitably engineered environments in related systems.
In addition to first-order SO interactions, here we also analyze in detail second-order effects, in particular,
their impact on the ground-state $g$ factors and on the structure of the on-site excitation spectrum.

\section{Ground-state $\mathbf{g}$~factors, experiment vs {\it ab initio} computations}

Within the family of layered Cu-oxide compounds, the type of $g$-factor anisotropy featured by Cu centers
in \ICVO, $g_{ab}\!>\!g_c$\,, is unusual since for the large majority of cuprates $g_{ab}\!<\!g_c$ is
commonly found ($g_{ab}$ and $g_{c}$ are here the in-plane and out-of-plane components of the ${\bf g}$
tensor, respectively).
This aspect was first noticed in Ref.~\cite{Kataev2004}, confirmed in Ref.~\cite{Okubo2011}, and attributed
to a less common ground-state configuration having one hole in the $z^2$ Cu 3$d$ orbital.
%
%
%
To better substantiate these earlier findings and interpretations, we perform in this work detailed
electronic-structure calculations for this material.
To illustrate and further document the experimental basis and to facilitate the discussion of the computational
results we first present additional ESR data measured on a powder sample of \ICVO, synthesized and thoroughly
characterized in Ref.~\cite{Moeller2008}.
The powder was mixed with an epoxy resin and the mixture was hardened in a magnetic field of several Tesla.
As a result, the so-prepared sample acquired a well-defined anisotropy axis (referred to hereafter as
orientation $o$ axis) \cite{Kataev2004} which according to x-ray diffraction measurements is perpendicular
to the crystallographic $c$ axis.

\begin{figure}[b]
\includegraphics[width=0.9\columnwidth]{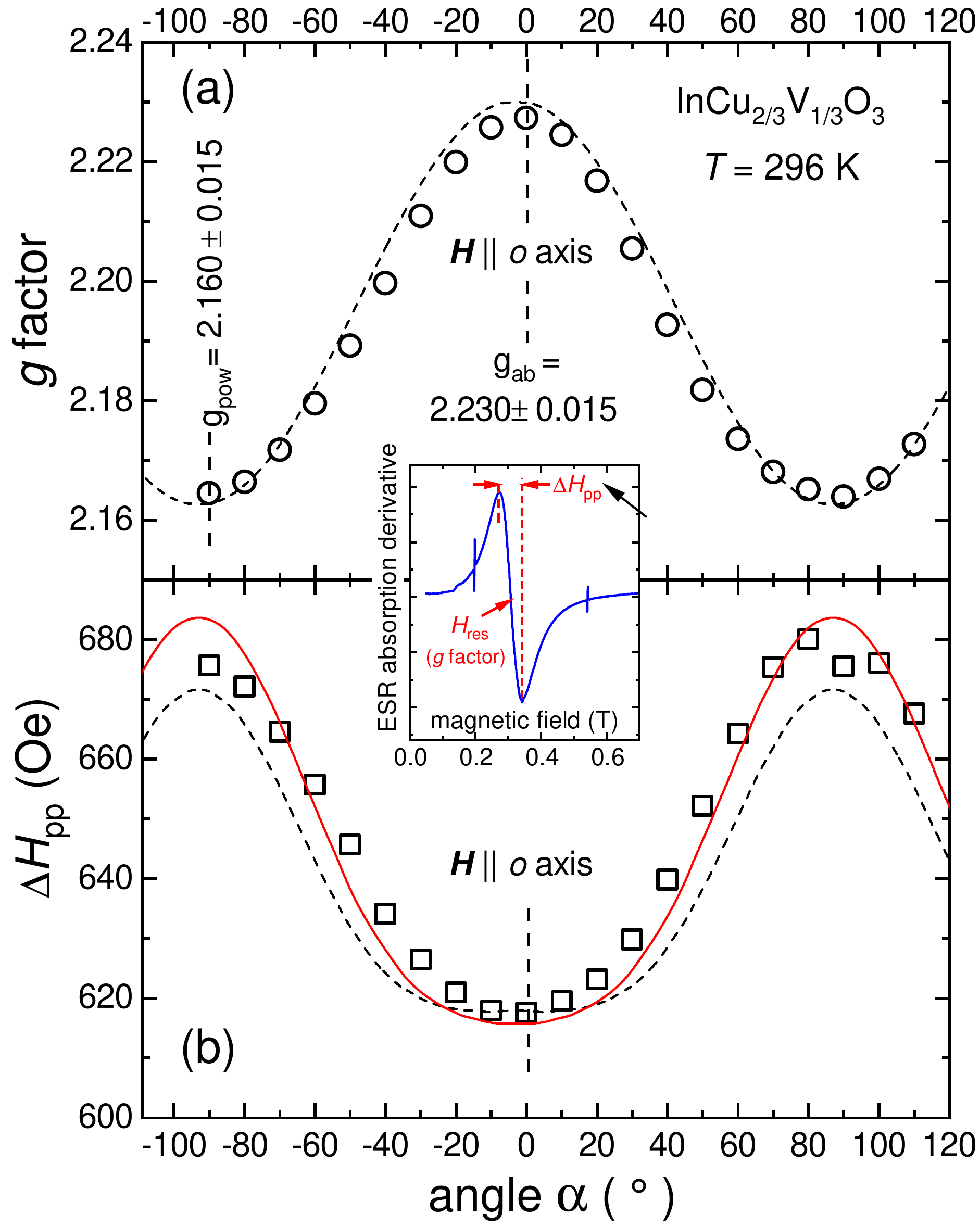}
\caption{
Angular dependence of the $g$~factor (a) and of the peak-to-peak ESR linewidth $\Delta H_{\rm pp}$
(b) of the oriented powder sample of \ICVO\ measured at room temperature at the X-band frequency
$\nu\!=\!9.58$\;GHz (symbols). 
The inset shows a typical ESR signal (field derivative of the microwave absorption).
Dashed and solid curves are the results of
numerical modeling (see
main
text).
}
\label{Xband}
\end{figure}

The dependence of the $g$~factor $g = h\nu/\mu_{\rm B}\mu_{0}H_{\rm res}$ and of the ESR peak-to-peak
linewidth $\Delta H_{\rm pp}$ on the angle $\alpha$ which the applied magnetic field ${\bf H}$ makes
with the $o$~axis are shown in Fig.~\ref{Xband}, as measured with a commercial X-band Bruker spectrometer
at a frequency $\nu=9.59$ GHz and room temperature.
$h$, $\mu_{\rm B}$, and $\mu_{0}$ are here the Planck constant, the Bohr magneton, and the vacuum
permeability, respectively.
The parameters of the ESR signal $H_{\rm res}$ (resonance field) and $\Delta H_{\rm pp}$ are defined
in the inset of Fig.\;\ref{Xband}.
The $g$~factor for $\alpha = 0$ corresponds to the single-crystalline in-plane $g$~factor $g_{ab}$,
whereas for $\alpha = \pm 90^\circ$ it yields the powder averaged $g$~factor $g_{\rm pow}$.
To model the experimental dependence of $g$ and $\Delta H_{\rm pp}$ on $\alpha$ we performed numerical
simulations of the ESR signal which account for the $\alpha$-dependent averaging of the contributions
from the individual crystallites in the sample.
The $g(\alpha)$ dependence can be very well reproduced, yielding the single-crystalline $g$~values
$g_{c} = 2.020 \pm 0.015$ and $g_{ab} = 2.230 \pm 0.015$ [dashed line in Fig.\;\ref{Xband}(a)]. 
The $g$-factor anisotropy should give rise to a dependence of the measured peak-to-peak linewidth
$\Delta H_{\rm pp}$ on the angle $\alpha$.
For $\alpha\!=\!0$, the ESR signal has the width corresponding to the single-crystalline line
$\Delta H$, $\Delta H_{\rm pp}(\alpha\!=\!0)\!=\!\Delta H$.
For $\alpha\!=\!\pm 90^\circ$, the ESR linewidth $\Delta H_{\rm pp}(\alpha\!=\!\pm 90^\circ)$ corresponds
to that of the powder sample and should be larger because of the distribution of the resonance fields
of the individual crystallites arising from the $g$-factor anisotropy,
$\Delta H_{\rm pp}(\alpha\!=\!\pm 90^\circ)\!>\!\Delta H$.
However, the respective $\Delta H_{\rm pp}(\alpha)$ modeled curve noticeably deviates from experiment
[dashed line in Fig.\;\ref{Xband}(b)], indicating that the $\Delta H_{\rm pp}(\alpha)$ dependence is
not entirely due to averaging signals with anisotropic $g$~factors.
Therefore, in addition, one could also consider a possible angular dependence of the width of individual
single-crystalline ESR lines of the form $\Delta H  = A + B\cdot(1+\cos^2\theta)$, where $\theta$ denotes
the angle between ${\bf H}$ and the $c$~axis.
Here, the first term is the linewidth in the absence of spin correlations at $T\rightarrow \infty$ and
the second term is the angular dependent contribution due to the three-dimensional (3D) antiferromagnetic
correlations at finite temperature \cite{Kubo1954,Huber1972,Richards1974,Benner1990,Zeisner2019}.
Indeed, accounting for this additional contribution in the simulation, with $A=571$\;Oe and $B=45$\;Oe,
significantly improves the agreement with the experiment [solid line in Fig.\;\ref{Xband}(b)].
The 3D-like behavior of \ICVO\ at elevated temperature is not surprising since the in-plane spin
correlations turning this compound into a 2D-$XY$ magnet begin to grow at temperatures $\lesssim 50$\;K
\cite{Iakovleva2019}.

\begin{figure}[t]
\includegraphics[width=0.9\columnwidth]{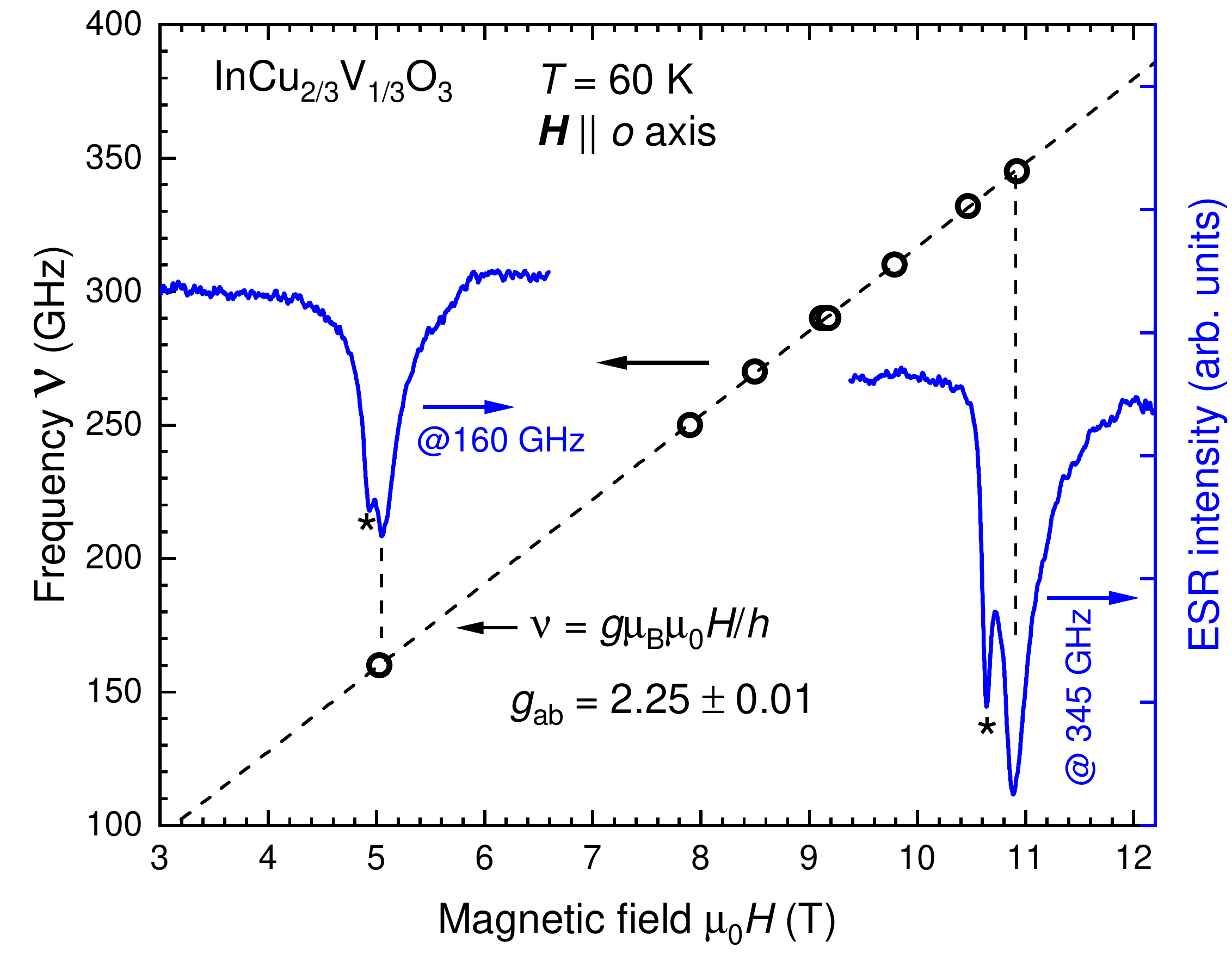}
\caption{
Frequency $\nu$ versus magnetic field $H$ dependence of the ESR signal of the oriented powder sample of
\ICVO\ measured at $T = 60$\;K with $\mathbf{H}\parallel o$~axis (open circles).
A corresponding fit is shown as dashed line.
Exemplary ESR signals are shown for frequencies of 160 and 345\;GHz.
The asterisk labels a small additional peak presumably due to Cu$^{2+}$ spins at defect sites.
}
\label{HFESR}
\end{figure}

Finally, a more accurate value can be obtained for $g_{ab}$ from the frequency $\nu$ versus field $H$
dependence of the ESR signal, which we measured using the home-made high-field ESR setup described
in \cite{Yehia2010} at a lower temperature $T\!=\!60$\;K (Fig.~\ref{HFESR}).
A fit of the data to the resonance condition $\nu = g_{\rm ab}\mu_{\rm B}\mu_{0}H/h$ yields
$g_{ab} = 2.25 \pm 0.01$.
With $g_{\rm pow}=2.23$, one obtains from the relation  $g_{\rm pow} = (1/3)g_{\rm c} +(2/3)g_{ab}$
\cite{Kataev2004} the $c$-axis $g$ factor $g_{c} = 1.99$.
The resulting ${\bf g}$ tensor $(g_{ab}, g_{c}) = (2.25, 1.99)$ is, within error bars of $\pm$5--10\%,
fully consistent with values obtained for other samples of \ICVO\ from various sources and for batches
earlier studied in Refs.~\cite{Kataev2004,Okubo2011}, providing altogether a solid reference for
theoretical analysis.

Two complementary computational techniques were employed for addressing the Cu
3$d$-shell electronic structure in \ICVO\ and peculiar features such as the
$g_{ab}\!>\!g_c$ anisotropy of the $g$~factor.  In a first step we determined the
optimal atomic positions in the frame of density-functional calculations with
periodic boundary conditions.
We used to this end the full-potential local-orbital code \textsc{fplo},
version 18~\cite{oj:FPLO}.  For nonmagnetic scalar-relativistic as well as
fully relativistic computations, we utilized the generalized gradient approximation
(GGA)~\cite{oj:PBE96}.  A $k$-mesh of 14$\times$14$\times$7 points was employed
in either case.  Typical for undoped cuprates, GGA yields a spurious metallic
state in \ICVO.
%
This well-known artifact stems from underestimating electronic correlations
with conventional functionals such as GGA.
The insulating state can be restored in DFT++ approaches, where interactions are considered explicitly on the level of (extended) multi-orbital Hubbard models~\cite{oj:lichtenstein98}. Nevertheless, the non-interacting part of such models is adopted from conventional DFT calculations. Hence, the respective model parameters, transfer integrals and crystal-field splittings, are not affected by shortcomings of the GGA.


\ICVO\ is a quasi-2D material whose honeycomb planes are only weakly coupled to
each other.  All samples synthesized so far feature a sizable number of
stacking faults, impeding a complete characterization of the crystal
structure~\cite{Moeller2008}.
Hence, the first step of our computational analysis
was
the
construction of a 3D structural model, based on the
experimentally available information: lattice constants and the internal
coordinates of a single
honeycomb
plane with Cu and V occupying the same crystallographic
site~\cite{Moeller2008}.  As explained in Ref.~\cite{Iakovleva2019}, a minimal
3D structural model for \ICVO\ employs the orthorhombic space group
$Cmcm$\,(63), which is compatible with
a
fully ordered honeycomb lattice of
Cu atoms and at the same time supports the simplest type of stacking of the
honeycomb planes.
The respective atomic coordinates are provided in Table~\ref{tab:str}.
The Cu-site point group 
symmetry is in this space group
$C_s$ instead of $D_{3h}$.

\begin{table}[tb]
\caption{\label{tab:str}
Atomic sites, Wyckoff positions (Wyck.\;pos.), and internal coordinates of the two different crystal
structures used in this study.
The space group is $Cmcm$ (63), with $a$\,=\,10.0527, $b$\,=\,5.80393, $c$\,=\,11.9012\,\AA.
In the 
computationally
optimized structure, the O positions were relaxed within the antiferromagnetic (N\'eel)
structure to yield the lowest total energy at the GGA+$U$ level.}
\begin{ruledtabular}
\begin{tabular}{ccp{.5em}rrrp{.5em}rrr}
\multirow{2}{*}{Site} & \multicolumn{1}{c}{Wyck.} && \multicolumn{3}{c}{Identical O$_5$ cages}
                                                  && \multicolumn{3}{c}{Optimized O positions} \\
& \multicolumn{1}{c}{pos.} && $x/a$ & $y/b$ & $z/c$ && $x/a$ & $y/b$ & $z/c$ \\ \hline
Cu  & $8g$ &&$\nicefrac13$&$\nicefrac13$& 0.25    &\\
V   & $4c$ && 0       &$\nicefrac13$& 0.25    &\\
In1 & $4a$ && 0       & 0       & 0       &\\
In2 & $8e$ &&$\nicefrac23$& 0       & 0       &\\
O1  & $4c$ && 0       & 0       & 0.25    && 0       & 0.03400 & 0.25    \\
O2  & $8g$ &&$\nicefrac13$& 0       & 0.25    && 0.35045 & 0.98309 & 0.25    \\
O3  & $8f$ && 0       &$\nicefrac23$& 0.5864 && 0       & 0.66708 & 0.58500 \\
O4  &$16h$ &&$\nicefrac13$&$\nicefrac13$& 0.0864 && 0.33313 & 0.33359 & 0.09067 \\
\end{tabular}
\end{ruledtabular}
\end{table}

The parent structural model~\cite{Moeller2008} does not distinguish between Cu and V,
featuring
identical O$_5$ cages around Cu and V atoms.
However, the honeycomb planes in \ICVO\ may undergo a breathing distortion, giving rise to unequal
Cu-O and V-O bond lengths.
To address this possibility, we performed a structural relaxation of O positions within the space
group $P2_1/m$, supporting N\'eel antiferromagnetic order.
To avoid inaccuracies stemming from the spurious metallic ground state, the optimization was carried
out with respect to the GGA+$U$ energy. Following earlier studies on low-dimensional cuprates~\cite{oj:ahmed15, oj:nawa17}, we use a Coulomb repulsion parameter $U$\,=\,8.5\,eV, a Hund's exchange $J$\,=\,1\,eV, and the fully localized limit for the double-counting correction. 
%
We find this way large differences between the in-plane Cu-O and V-O bond lengths, 2.04 vs 1.74 \AA , 
consistent with considerations concerning the difference between the formal oxidation states of
copper and vanadium in this system \cite{Moeller2008}.
Also,
the computationally derived lattice parameters imply
significantly longer in-plane Cu-O links as compared to the apical Cu-O bonds
(2.04 vs 1.90 \AA ).

Using crystallographic data as listed in Table\;\ref{tab:str}, we further performed many-body quantum
chemical calculations \cite{Helgaker2000} on a finite atomic fragment having a CuO$_5$ unit as central
region (all relevant computational details are provided in Appendix).
To describe configurational mixing effects mediated by SO interactions and related $g$-factor anisotropies
we relied on spin-orbit multiconfiguration and multireference numerical schemes as implemented in the
quantum chemical package {\sc molpro} \cite{molpro12} by Berning {\it et al.}\;\cite{SOC_molpro}.
An active space defined by the Cu 3$d$ orbitals (nine electrons in five orbitals) was used to this end
in preliminary multiconfiguration computations \cite{Helgaker2000}.
All possible states associated with this 3$d$-shell filling were considered, which in quantum chemical
terminology is referred to as complete-active-space self-consistent-field (CASSCF) methodology
\footnote{
However, for the 3$d^9$ manifold and sufficiently high symmetry, this essentially corresponds to
single-configuration restricted open-shell Hartree-Fock (ROHF) \cite{Helgaker2000}.
}.
The $z^2$ character of the Cu 3$d$ hole proposed on the basis of the measured $g_{ab}\!>\!g_c$ structure of
the $g$ factors \cite{Kataev2004} is also confirmed in the CASSCF calculation. 
For making direct connection with the experimental data, $g$ factors were also computed:
using CASSCF wave-functions, related angular-momentum and SO matrix elements \cite{SOC_molpro}, and the
methodology described in Ref.\;\cite{Bogdanov15}, we arrive to $g_{ab}\!=\!2.23$ and $g_c\!=\!1.96$;
on the basis of multireference configuration-interaction (MRCI) wave-functions built by additionally
considering single and double excitations \cite{Knowles92} out of the Cu 3$d$ and O 2$p$ orbitals of the
CuO$_5$ polyhedron, we obtain $g_{ab}\!=\!2.21$ and $g_c\!=\!1.97$.
These CASSCF and MRCI values are within 98--99$\%$ of the estimates derived from the ESR data and presented
above.
For a more detailed picture on the Cu 3$d$-shell electronic structure in this compound, we analyze in the
following the Cu $d$-$d$ excitations since the $g$-factor values depend on the relative energies of the
excited states.

\begin{table}[!b]
\caption{
Cu$^{2+}$ 3$d^9$ multiplet structure (relative energies in eV) using atomic positions as obtained by
density-functional optimization.
The MRCI values include Davidson corrections \cite{Helgaker2000}.
The character of the SO wave-functions is also specified.
}
\begin{tabular}{c c c l}
\hline
\hline\\[-0.30cm]
$3d^9$ states  &CASSCF &MRCI  &SO-MRCI\\
\hline
\\[-0.20cm]
$^2\!A_1'$     &0      &0     &0 \ \ \ \ (99$\%$ $A_1'$ character)\\[0.02cm]

$^2\!E''$      &1.03   &1.20  &1.12 (89$\%$ $E''$, 11$\%$ $E'$)\\[0.02cm]
               &       &      &1.21 (99$\%$ $E'$)              \\[0.02cm]

$^2\!E'$       &1.12   &1.30  &1.28 (99$\%$ $E''$)             \\[0.02cm]
               &       &      &1.46 (89$\%$ $E'$, 11$\%$ $E''$)\\
\hline
\hline
\end{tabular}
\label{DFT_latt}
\end{table}

\section{Excited states}


CASSCF, MRCI, and spin-orbit MRCI (SO-MRCI) results for the Cu$^{2+}$ multiplet structure using atomic
positions as determined by density-functional lattice optimization are listed in Table\;\ref{DFT_latt}.
Notations corresponding to $D_{3h}$ point-group symmetry are employed, where the Cu $z^2$, ($xy$,\,$x^2\!-\!y^2$),
and ($xz$,\,$yz$) orbitals belong to the irreducible representations $A_1'$, $E'$, and $E''$, respectively.
It is found that the two sets of doubly degenerate levels lie rather close in energy and that the MRCI
treatment brings corrections of 0.1--0.2 eV to the CASSCF crystal-field splittings.
Rectifications in this range were previously computed at the MRCI level in other copper oxide compounds
\cite{CuO2_dd_hozoi_11,CuO_dd_huang_11} and mainly originate from L\,2$p$\,--\,TM\,3$d$ charge-transfer-type
correlation effects.
SO interactions lift the degeneracy of the $E'$ and $E''$ states and additionally lead to some degree of
$E'\!-\!E''$ mixing.
This is an aspect that deserves attention:
since $A_1'$ to $E''$ dipole transitions are not allowed, the three-peak structure of the optical absorption
spectrum \cite{Moeller2008} can be then qualitatively understood on the basis of the fact that only three
of the four SO excited states have significant $E'$ character (see the right column in Table\;\ref{DFT_latt}).

The $E'\!-\!E''$ mixing is obviously the effect of second-order SO couplings.
Such physics was not
addressed in detail
in earlier analysis of the Cu $d$--$d$ excitation spectrum of \ICVO\ but
given the near-degeneracy of the $E'$ and $E''$ crystal-field-like states it plays a significant role.
To better illustrate this point, we performed SO calculations for each of the $E'$ and $E''$ terms separately.
The splittings induced by first-order SO interactions came out as 200 meV and 101 meV, respectively.
Taking as reference the MRCI relative energies provided in Table\;\ref{DFT_latt} (1.20 and 1.30 eV) and
neglecting second-order SO couplings, these first-order splittings translate into excitation energies of
1.15, 1.25 eV ($E''$ terms) and 1.20, 1.40 eV ($E'$ terms) for the SO excited states, different from the
results of the full SO-MRCI computation (1.12, 1.21, 1.28, and 1.46 eV).
Compared to earlier calculations based on the angular overlap model (AOM) \cite{Moeller2008}, our {\it 
ab initio} quantum chemical results differ therefore in two aspects:
a different sequence of the $E'$ and $E''$ terms in the absence of first-order SO interactions
($E(^2\!E')\!>\!E(^2\!E'')$ in Table\;\ref{DFT_latt} while $E(^2\!E')\!<\!E(^2\!E'')$ in the AOM
model \cite{Moeller2008}) and significantly stronger first-order SO interactions for the $E''$
states (split in first order by 101 meV here, see above, and by 62 meV (500 cm$^{-1}$) in the AOM
model \cite{Moeller2008}).


As concerns a direct comparison between the relative energies of the SO states with significant $E'$
character listed in Table\;\ref{DFT_latt} and the positions of the peaks in the optical absorption
spectrum \cite{Moeller2008}, it is seen that the agreement is not perfect: 1.12, 1.21, and 1.46 eV
vs 1.18, 1.35, and 1.60 eV (9500, 10900, and 12900 cm$^{-1}$ in Ref.\;\cite{Moeller2008}).
Deviations of $\sim$0.1 eV between MRCI excitation energies and experimental peak positions, with
MRCI constantly underestimating experimental values, are however common for Cu$^{2+}$ 3$d^9$ oxide
compounds \cite{CuO2_dd_hozoi_11,CuO_dd_huang_11}.

For completeness, we additionally provide in Table\;\ref{idealized_latt} on-site excitation energies
computed on the basis of the more idealized crystal structure
with identical MO$_5$ cages.
The crystal-field $a_1'\!-\!e'$ and $a_1'\!-\!e''$ splittings (which we denote as $\delta'$ and $\delta''$,
respectively) are different in this case, with $\delta'\!<\!\delta''$.
A reversed sequence of the $e'$ and $e''$ levels in the two different crystal structures is also found
by density-functional theory, by Wannier projection of the GGA bands onto the five 3$d$ orbitals at
a Cu site, as discussed in Appendix.
Most importantly, with SO couplings accounted for, the calculated excitation energies are far from the peak
positions found in optics (see  Table\;\ref{idealized_latt}).
In particular, `pairs' of nearly-degenerate SO states are obtained computationally for the more idealized
crystal structure, different from the three-peak structure observed experimentally \cite{Moeller2008}.
A more consistent picture in modeling experimental data was also found in the AOM frame \cite{Moeller2008}
when elongated in-plane Cu-O bonds are employed.

\begin{table}[!t]
\caption{
Results for the Cu$^{2+}$ 3$d^9$ multiplet structure (relative energies in eV) using the idealized
lattice configuration proposed in Ref.~\;\cite{Moeller2008}.
The MRCI values include Davidson corrections \cite{Helgaker2000}.
The character of the SO wave-functions is also provided.
}
\begin{tabular}{c c c l}
\hline
\hline\\[-0.30cm]
$3d^9$ states  &CASSCF &MRCI   &SO-MRCI\\
\hline
\\[-0.20cm]
$^2\!A_1'$     &0      &0      &0 \ \ \ \ (99$\%$ $A_1'$ character)\\[0.02cm]

$^2\!E'$       &0.98   &1.17   &1.08 (99$\%$ $E'$)              \\[0.02cm]
               &       &       &1.12 (72$\%$ $E''$, 28$\%$ $E'$)\\[0.02cm]

$^2\!E''$      &1.03   &1.22   &1.30 (99$\%$ $E''$)             \\[0.02cm]
               &       &       &1.34 (72$\%$ $E'$, 28$\%$ $E''$)\\
\hline
\hline
\end{tabular}
\label{idealized_latt}
\end{table}

\section{Conclusions}

In sum, a detailed analysis of the Cu $d$-shell electronic structure in the quasi-2D honeycomb compound
\ICVO\ is performed, with focus on the interplay of electron correlations, spin-orbit
couplings, and structural effects in this material.
{\it Ab initio} computational data are compared to the outcome of ESR experiments and of previous
optical absorption measurements \cite{Moeller2008}.
The three-peak structure of the optical spectra is reproduced by many-body quantum chemical calculations
only for atomic positions obtained from a prior lattice optimization relying on density-functional theory
and not for a more idealized crystal structure
with identical MO$_5$ cages.
It seems to be related with the fact that $E'\!-\!E''$ mixing occurs for only two of the spin-orbit excited
states.
To describe electron correlation effects, we relied on multireference configuration-interaction calculations
with single and double excitations out of the copper 3$d$ and oxygen 2$p$ shells within a given CuO$_5$ unit.
Although the correlation-induced corrections obtained this way to the $d$--$d$ excitation energies are
sizable, 0.1--0.2 eV, deviations in the range of 0.1 eV still remain as compared to experimental peak
positions.
On the other hand, the computed ground-state $g$~factors are in excellent agreement with values derived
from all available ESR data, within 98--99$\%$.
The peculiar coordination generates strong axial anisotropy at copper sites in \ICVO,
reminiscent to some exent of, e.\,g., linearly coordinated transition-metal ions \cite{fe_d7_Zadrozny_2013,
fe_d7_Jesche_2014,fe_d7_xu_2017,fe_d7_Jesche_2015,fe_d7_Antropov_2014}.
Adjusting the sequence of crystal-field levels through adequate design of the chemical/electrostatic
environment \cite{Bogdanov15,f13_ziba_19} such that the $z^2$ orbital is filled leaves room for interesting
single-ion magnetic properties even for the 3$d^9$ electron configuration, as pointed out in Refs.
\cite{fe_d7_Jesche_2015,fe_d7_Antropov_2014}.

%
%

\section*{Acknowledgements}

We thank A.~Jesche for useful discussions, U.~Nitzsche for technical assistance, and
A.~M\"oller for providing a sample of \ICVO\ for the ESR experiment and for important comments to
the manuscript.
L.\,H. acknowledges financial support from the German Research Foundation (Deutsche Forschungsgemeinschaft,
DFG), grant HO-4427/3.
O.\,J. was supported by the Leibniz Association through the Leibniz Competition.

\appendix
\section{Computational details}

To evaluate crystal-field parameters and the SO coupling constant $\zeta$
within the GGA, we performed Wannier projections of the GGA bands onto the five
3$d$ orbitals at a Cu site.  As these orbitals hybridize with oxygen $2p$ and
vanadium $3d$ orbitals, the resulting hoppings sensibly depend on the energy
window used for projections.  Moreover, the Fourier-transformed Wannier
functions show deviations from the GGA bands at the bottom of the valence band,
especially for the experimental structure.  However, we are interested
exclusively in the local terms, and these show only a small (on the order of
several meV) dependence on the energy window.  Using parameters obtained from
such a procedure, we construct the local Hamiltonian $H_0$. Corresponding
eigenvalues are provided for both crystal structures in Table~\ref{tab:wf}.

For both lattice configurations, $H_0$ features two pairs of nearly degenerate
eigenstates: $\lambda_1\simeq\lambda_2$ and $\lambda_3\simeq\lambda_4$.
But the gaps between them, $\Delta_1\!=\!\lambda_3\!-\!\lambda_2$ and
$\Delta_2\!=\!\lambda_5\!-\!\lambda_4$, differ significantly in \ICVO:
$\Delta_1\!\simeq\!0.30$\,(0.19)\,eV is substantially smaller than
$\Delta_2\!\simeq\!1.06$\,(1.10)\,eV in the idealized (optimized) structure.
As concerns the sequence of the two groups of nearly degenerate states, we found that
in the more idealized crystal structure $\lambda_1$ has dominant (84\%)
$xz$ character while $\lambda_2$ is dominated (84\%) by the $yz$ orbital.
Similarly, $\lambda_3$ and $\lambda_4$ pertain to the $xy$ and $x^2\!-\!y^2$ orbitals,
respectively.  But in the optimized structure, the order is reversed: the
lower-lying eigenstates in each group, $\lambda_1$ and $\lambda_3$, are
dominated by $yz$ (94\%) and $x^2\!-\!y^2$ (61\%) orbitals.

\begin{table}[t!]
\caption{\label{tab:wf}
Eigenvalues $\lambda_n$ of the local Hamiltonian $H_0$ parameterized using 
Wannier projection.
In the fully relativistic calculations (GGA+SOC), all eigenvalues are doubly degenerate.
All values are given in eV with respect to the Fermi level.
}
\begin{ruledtabular}
\begin{tabular}{rcccc}
Eigenvalues & \multicolumn{2}{c}{Identical O$_5$ cages}
            & \multicolumn{2}{c}{Optimized O positions} \\
of $H_0$ & GGA & GGA+SOC & GGA & GGA+SOC \\ \hline
$\lambda_1$ & $-1.61$ &  $-1.65$  &  $-1.44$ &  $-1.48$ \\
$\lambda_2$ & $-1.60$ &  $-1.60$  &  $-1.43$ &  $-1.46$ \\
$\lambda_3$ & $-1.31$ &  $-1.37$  &  $-1.25$ &  $-1.29$ \\
$\lambda_4$ & $-1.31$ &  $-1.24$  &  $-1.25$ &  $-1.17$ \\
$\lambda_5$ & $-0.26$ &  $-0.26$  &  $-0.15$ &  $-0.15$ \\
\end{tabular}
\end{ruledtabular}
\end{table}

Finally, we applied a fully relativistic GGA treatment and performed Wannier
projections to estimate the SO coupling constant $\zeta$.
The respective eigenvalues are provided in Table~\ref{tab:wf} next to those
for the scalar-relativistic case
\footnote{Note that the fully relativistic calculations are nonmagnetic and
all eigenvalues are doubly degenerate.}.
The SO coupling $\zeta$ can be then estimated by averaging over two offdiagonal
elements of $H_0$:
\begin{equation}
\label{eq:zeta}
\zeta = -\frac{1}{2\sqrt{3}}\left(i\langle{}\phi_{z^2}^{\uparrow}|H_0|\phi_{yz}^{\downarrow}\rangle + \langle{}\phi_{z^2}^{\uparrow}|H_0|\phi_{xz}^{\downarrow}\rangle\right),
\end{equation}
where $\phi_{\alpha}^{\sigma}$ is the basis (Wannier) function of orbital
$\alpha$ and spin $\sigma$. 
We obtain this way $\zeta$\,=\,68(60)\,meV for the idealized (optimized) structure.

As concerns the quantum chemical calculations, the actual cluster considered in the computations
consists of a central CuO$_5$ unit along with three adjacent VO$_5$ polyhedra, three Cu nearest
neighbors, and six adjacent In ions.
The remaining part of the extended crystalline surroundings was modeled as an effective electrostatic
field \cite{PCembedd_klintenberg_00}.
For the central CuO$_5$ unit, all-electron Douglas-Kroll basis sets of triple-$\zeta$ quality with
polarization functions were used.
For farther oxygen ligands in the cluster and for the nearby V ions, we applied basis sets of
double-$\zeta$ quality.
The results discussed in the main text were obtained by modeling the Cu$^{2+}$ nearest neighbors as
closed-shell Zn$^{2+}$ total-ion potentials (TIP's), an approximation also employed in earlier quantum
chemical investigations \cite{CuO2_dd_hozoi_11,CuO_dd_huang_11,Nikolay_RIXS}.
Test CASSCF calculations in which the three adjacent Cu species are represented as 2+ $S\!=\!1/2$ 3$d^9$ ions
provide central-site crystal-field splittings that agree within 0.01 eV with those computed with Zn$^{2+}$
TIP's.
We used the latter for all subsequent computations since this makes the spin-orbit calculations and
related analysis much less complex.
TIP's were also applied for the nearby In$^{3+}$ ions.

\bibliography{ESR_jul14,InCuVO_oj_aug04}

\end{document}